\documentstyle[12pt,epsfig]{article}
\def\bxll{B \rightarrow X \, \l^+ \, \l^-}
\def\bsg{B \rightarrow X_s \, \gamma}

\def\bsmm{B^0_q \rightarrow l^+ l^-}

\def\beq{\begin{eqnarray}}
\def\eeq{\end{eqnarray}}
\def\nnb{\nonumber}
\def\aB0{\big| B^0 \big>}
\def\bB0{\big| \bar{B}^0 \big>}
\begin{document}

\begin{center}
{\LARGE\bf
Search for new physics via CP violation in $B_{d,s} \to l^+
l^-$ }
\vspace*{0.5cm}

       {\bf Chao-Shang HUANG}\footnote{E-mail : csh@itp.ac.cn},
        {\bf LIAO Wei }\footnote{E-mail: liaow@itp.ac.cn}
        \vspace{0.5cm}

        Institute of Theoretical Physics,
         Academia Sinica, 100080 Beijing, China \\
\vspace{0.5cm}
\begin{abstract}
It is shown that in the approximation of $|\frac{q}{p}|$=1 the CP violation 
in $B^0_{d,s} \to l^+ l^-$ decays vanishes in SM. In a 2HDM with CP 
violating phases and MSSM the CP asymmetries depend on the parameters of
models and can be as large as $40\%$ for $B^0_d$ and $3\%$ for $B^0_s$.
An observation of CP asymmetry in the decays would unambiguously signal
the existence of new physics.

\end{abstract}
\end{center}
\noindent
\vfill
\vskip 0.5cm
\centerline{{\sc Pacs} numbers: 11.30.E, 13.20.H, 12.60.F, 12.60.J}
\renewcommand{\thefootnote}{\arabic{footnote}}
\setcounter{footnote}{0}
\vskip 0.5cm

\indent
The flavor changing neutral current process, $B_{d,s} \to l^+ l^-$
(l=$\mu,\tau$), 
has attracted a lot attention since it is very sensitive to the structure of
SM and potential new physics beyond SM and was shown to be powerful to shed
light on the existence of new physics before possible new particles are
produced at colliders\cite{hnv,hy,hlyz}. In a very large region of
parameter space supersymmetric(SUSY) contributions were shown to be easy
to overwhelm the SM contribution\cite{hy,hlyz,cs} and even reach, e.g., for l=$\mu$, the experimental 
upper bound\cite{data}
\beq
B_r (B_d \rightarrow \mu^{+}\mu^-) < 6.8 \times 10^{-7} \ \ (CL=90\% ) \nnb \\
B_r (B_s \rightarrow \mu^{+}\mu^-) < 2.0 \times 10^{-6} \ \ (CL=90\% ). 
\label{limit}
\eeq
In other words measuring the branching ratio of $B_{d,s} \to l^+ l^-$ can give 
stringent constraints on the parameter space of new models beyond SM, especially
for that of the minimal supersymmetric standard model (MSSM) because of the 
tan$^3\beta$ dependence of SUSY contributions in some large tan$\beta$ regions
of the parameter space\cite{hy,hlyz}.
Comparing with hadronic decays of B mesons, this process is very clean
and the only nonperturbative quantity involved is the decay constant
that can be calculated by using lattice, QCD sum rules etc. 

The results on CP violation in $B_d$ - $\bar B_d$
mixing have been reported by the BaBar and Belle Collaborations\cite{Osaka}
in the ICHEP2000 Conference, which are consistent with the world average\cite{cdf} . More experiments on B physics have been planned
in the present and future B factories \cite{bf}. In the letter we study CP violation
in $B_{d,s} \to l^+ l^-$ (l=$\mu, \tau$), which might be measured in the near future. 

Obviously for the process $B_{d,s} \to l^+ l^-$  there are no direct CP
violations since there are no strong phases in the decay amplitude\cite{hnv,hy,hlyz}. 
But it is well known that CP violating effects can survive after taking into account the
mixing of the neutral mesons, $B^0$ and $\bar{B}^0$, in the absence of the
strong phases. We will give
a model-independent description for the CP violating effects of the process
induced by mixing of $B^0$ and $\bar{B}^0$ and
analyze them in SM and new models, a two Higgs doublet model(2HDM) with CP violating
phases and MSSM. 

We need to know what kind of CP violating observables can be defined
in the process. At first, direct CP violation, as noted above, is 
absent in this process. T-odd projection of polarization is a kind of
useful tool to probe the CP violating effects, for example, in $\bxll$
\cite{ks,hl,hz}. However for the process we are discussing here, we have
actually only one independent momentum and one independent spin
which can be chosen as those of $l^-$, so no T-odd projections can
be defined. Unlike the case generally discussed for hadronic final states,
for example, that in Ref.~\cite{gron}, the detected final states of
$l^+$ and $l^-$ of this process in experiments are basically 
two asymptotic energy-momentum eigenstates which are not CP eigenstates. Considering
for instance $B^0$ decays to $l^+ l^-$ in the rest frame of $B^0$,
due to the energy-momentum conservation we denote the four-momenta of
$l^-$ and $l^+$ as $p=(E,\vec{p})$ and $\bar{p}=(E,-\vec{p})$.
Then the angular momentum conservation tells us that $l^+_L l^-_R$
and $l^+_R l^-_L$ final states are forbidden. Hence we are left
with a pair of CP conjugated final states, $l^+_L l^-_L$ and
$l^+_R l^-_R$ and two couple of corresponding CP conjugated 
processes. Therefore, we may define the time dependent CP asymmetries as
\beq
A^1_{CP}(t)&=&
\frac{\Gamma(B^0_{phys}(t) \rightarrow l^+_L l^-_L)
-\Gamma(\bar{B}^0_{phys}(t) \rightarrow l^+_R l^-_R)}
{\Gamma(B^0_{phys}(t) \rightarrow l^+_L l^-_L)
+\Gamma(\bar{B}^0_{phys}(t) \rightarrow l^+_R l^-_R)}
\label{cpt1} \\
A^2_{CP}(t)&=&
\frac{~\Gamma(B^0_{phys}(t) \rightarrow l^+_R l^-_R)
-\Gamma(\bar{B}^0_{phys}(t) \rightarrow l^+_L l^-_L)}
{\Gamma(B^0_{phys}(t) \rightarrow l^+_R l^-_R)
+\Gamma(\bar{B}^0_{phys}(t) \rightarrow l^+_L l^-_L)}
\label{cpt2}
\eeq
Two corresponding time integrated CP asymmetries are
\beq
A^i_{CP}&=&
\frac{\int^\infty_0 dt ~\Gamma(B^0_{phys}(t) \rightarrow f_i)
-\int^\infty_0 dt ~\Gamma(\bar{B}^0_{phys}(t) \rightarrow \bar{f}_i)}
{\int^\infty_0 dt ~\Gamma(B^0_{phys}(t) \rightarrow f_i)
+\int^\infty_0 dt ~\Gamma(\bar{B}^0_{phys}(t) \rightarrow \bar{f}_i)}~~~~~~~i=1, 2
\label{cp1} 
\eeq
Where $f_{1,2}=l^+_{L,R}l^-_{L,R}$, $\bar{f}$ is the CP conjugated state of $f$.


The time evolutions of the initial pure 
$B^0$ and $\bar{B}^0$ states are given by\cite{75}
\beq
&& \big| B^0_{phys}(t) \big> = g_+(t)\aB0 + \frac{q}{p} g_-(t) \bB0, \nnb \\ 
&& \big| \bar{B}^0_{phys}(t) \big> = \frac{p}{q} g_-(t)\aB0 + g_+(t)\bB0. \label{b0t}
\eeq
with $g_{\pm}(t)$ given by
\beq
&& g_+(t) = exp(-\frac{1}{2}\Gamma t-i m t) cos(\frac{\Delta m}{2} t), \nnb \\
&& g_-(t) = exp(-\frac{1}{2}\Gamma t-i m t) i sin(\frac{\Delta m}{2} t) \label{gt}
\eeq

 The absence of strong phases implies
\beq
|A_f|=|\bar{A}_{\bar{f}}|, ~~~~|A_{\bar{f}}|=|\bar{A}_f|
\label{axi}
\eeq
where $A_f(\bar{A}_f)= < f|  {\cal H}_{eff} | B^0 (\bar{B}^0)>$.
And the CPT invariance leads to
\beq
\frac{\bar{A}_f}{A_f} =  \bigg( \frac{A_{\bar{f}}}{\bar{A}_{\bar{f}}} \bigg)^*
\label{rate11} 
\eeq
For simplicity, define
\begin{equation}
\frac{\bar{A}_f}{A_f}=\rho e^{i\phi_f},~~~~\frac{q}{p}=x e^{i\phi_x}.\nnb
\end{equation}

From Eqs.(\ref{cpt1}), (\ref{cpt2}), (\ref{b0t}), (\ref{axi}), and (\ref{rate11}),
 it is straightforward to derive
\beq
r\equiv |\frac{A(\bar{B}^0(t)\rightarrow \bar{f})}{A(B^0(t)\rightarrow f)}|
= \frac{|1+x^{-1}\rho tan(\frac{\Delta m}{2} t)exp[i(-\phi_f-\phi_x+\frac{\pi}{2})]|}
{|1+x \rho tan(\frac{\Delta m}{2} t)exp[i(\phi_f+\phi_x+\frac{\pi}{2})]|}.
\eeq
Therefore, if
\beq
 x\neq 1
\label{x1}
\eeq
or
\beq
\phi_f+\phi_x\neq 0~~~~mod~~ 2n\pi,
\label{phi1}
\eeq
( or equivalently $Im (\frac{q}{p}\frac{\bar{A}_f}{A_f})\neq 0$, )
then $r\neq 1$, i.e., one has CP violation.

The effective Hamiltonian governing the process $B_{d,s} \to l^+ l^-$ 
has been given in Refs. \cite{hy,hlyz}. Using the effective Hamiltonian,
we obtain by a straightforward calculation \footnote {We have neglected the contributions, which is smaller 
than or equal to $10^{-3}$ of the leading term, from the penguin diagrams with c and u quarks in the loop. 
It is true for both $B_d$ and $B_s$ decays\cite{bur}. Therefore, although there are  weak phases from the c 
or u quark in the loop, in particular,
 for $B_d$, the effect on the decay phase induced by them is neglegiblly small.}
\beq
\frac{\bar{A}_{f_1}}{A_{f_1}} = 
 -\frac{\lambda_t}{\lambda_t^*} \frac{C_{Q1}\sqrt{1-4\hat{m}_l^2}
+ (C_{Q2}+2 \hat{m}_l C_{10})}{C^*_{Q1}\sqrt{1-4\hat{m}_l^2}
- (C^*_{Q2}+2 \hat{m}_l C^*_{10})},\label{rate}
\eeq
where $\lambda_t=V_{tb} V_{td}^*$ or $V_{tb} V_{ts}^*$, $\hat{m}_l =
m_l/m_{B^0}$ and $C_i$'s 
are understood as Wilson coefficients at $m_B$ scale\cite{wise,buras,mas,hy,hlyz}. Because $C_{Q_i}$'s are
proportional to $m_l$ and $C_{10}$ is independent of $m_l$ it follows from eq. (\ref{rate}) that CP asymmetry in $B_{d,s} 
\to l^+ l^-$ is independent of the mass of the lepton. That is, it is the same for l = electron, muon and tau.
 
In SM, one has\cite{nir}\footnote{Note that the phase convention between $B^0$ and $\bar{B}^0$ 
is fixed as ${\cal CP}|B^0> = - |\bar{B}^0>$ when deriving eqs. (\ref{rate}), (\ref{mix}).}
\beq
\frac{q}{p}= - \frac{M^*_{12}}{|M_{12}|}= - \frac{\lambda^*_t}{\lambda_t},
\label{mix}
\eeq 
up to the correction smaller than or equal to order of $10^{-2}$, $C_{10}$ 
is real, $C_{Q_2}=0$, and $C_{Q_1}$ is negligibly small. So it follows from
Eqs.(\ref{rate}), (\ref{mix}) that $x=1$ and $\phi_f+\phi_x$=0. Therefore,
there is no CP violation in SM. 
\footnote{One can check by combining Eqs. (\ref{mix}) and (\ref{rate})
that all freedoms of phase conventions are canceled out completely in
$\frac{q}{p}\frac{\bar{A}_{f_1}}{A_{f_1}}$, including the one between $B^0$ and
$\bar{B}^0$.  }
If  one includes the correction smaller than order of $10^{-2}$ to $x$=1\footnote{According to the box
diagram calculation in SM, the deviation of x from 1 is $\sim 10^{-3} (10^{-5})$ for $B_d (B_s)$\cite{nir}. So $10^{-2}$ is a
conservative estimate.} one will have  
CP violation of order of $10^{-3}$ for $B^0_d$ and 
$10^{-4}$ for $B^0_s$ which are unobservably small. 

In the models where Eq. (\ref{mix}) is valid, 
defining $\xi=\frac{\bar{A}_{f_1}}{A_{f_1}}$, $\bar{\xi}=\frac{q}{p}\xi$
and using Eqs. (\ref{rate11}), (\ref{mix}),
the time dependent CP asymmetries Eqs. (\ref{cpt1}) and (\ref{cpt2})
and time integrated CP asymmetries Eq. (\ref{cp1}) 
can be greatly simplified 
\beq
A^1_{CP}(t) &=& -\frac{sin(\Delta m t) Im(\bar{\xi})}{cos^2(\frac{1}{2}
\Delta m t)+ |\xi|^2 sin^2(\frac{1}{2}\Delta m t)} \label{CPT1} \\
A^2_{CP}(t) &=& -\frac{sin(\Delta m t) Im(\bar{\xi})}{|\xi|^2 cos^2
(\frac{1}{2}\Delta m t)+ sin^2(\frac{1}{2}\Delta m t)} \label{CPT2} \\
A^1_{CP} &=& 
 -\frac{2 Im(\bar{\xi}) X_q}{(2+X^2_q)+X^2_q|\xi|^2}
\label{CP1} \\
A^2_{CP} &=&
-\frac{2 Im(\bar{\xi}) X_q}{(2+X^2_q)|\xi|^2+X^2_q}
\label{CP2}
\eeq
where $X_q= \frac{\Delta m_q}{\Gamma}$($q=d,s$ for $B^0_d$ and $B^0_s$
respectively). As expected, they are nonzero in the presence of CP violating phases.

We have discussed the CP asymmetries assuming that $B^0$ or $\bar{B}^0$
mesons are tagged before the decay $\bsmm$($q=d,s$) happen \footnote{An analysis of
tagging has been carried out in Refs.\cite{ros,gron}.} Likewise one may
also define CP asymmetries of the opposite tagging order\cite{gron} which turn
out to be just of the opposite sign of those defined above, (\ref{CP1}) and (\ref{CP2}).
 (Eqs. (\ref{CPT1}) and (\ref{CPT2}) hold for either tagging order.)  The CP asymmetries not
requiring measurement of the time order as one may
naively imagine to define, however, turn out to be zero because of the
the relation Eq. (\ref{rate11}) and the approximation Eq. (\ref{mix}) we
have used in our discussions. 

From Eq. (\ref{CP1}) and  (\ref{CP2}) one can simply get the  maximal 
limit of the CP violating observables
\beq
|A^{1,2}_{CP}(B^0_q)|_{max}=\frac{1}{\sqrt{2+X_q^2}}
\label{max1}
\eeq
which is about $63\%$ for q=d and $5\%$ for q=s. For $B^0_s$ we know that $X_s$
is experimentally larger than $15.7$($90 \% ~CL$)\cite{data}, so
we can neglect the number 2 in the
formula and get
\beq
A^2_{CP}(B^0_s) \doteq
-\frac{2 Im(\bar{\xi}) X_s}{X^2_s|\xi_s|^2+X^2_s}
\doteq A^1_{CP}(B^0_s). \nnb
\eeq
The situation is clearly quite different for $B^0_d$ because $X_d$ is 
just about $0.73$.  The two CP asymmetries for $B^0_d$ do not
exhibit strong correlation. 

In Fig. 1 the correlation between the CP asymmetries of $B_s^0$ and $B^0_d$ is
 plotted scanning the parameter space of
$C_{Q_1}$ and $C_{Q_2}$ with $|C_{Q_i}| \ge 0.1$ ($i=1,2$).
The points in the figure are plotted satisfying the constraints Eq. (\ref{limit}). One sees 
that they do not exhibit strong correlation in the parameter space which is actually
implied by the fact that in the most of the parameter space, $|\xi_s|^2$ 
(of order one) is not important at all because of the very large $X_s^2$,
while $|\xi_d|^2$ would compete with $X^2_d$ in
the formula Eq. (\ref{CP1}).

We now discuss CP violation of the process in a general 2HDM and MSSM. It has been shown 
\cite{ck} that the contributions to the mixing 
of $B^0$ and $\bar{B}^0$  from 2HDM or MSSM can be significant when the charged
Higgs boson mass and tan$\beta$ are small($m_{H^\pm}\le 200$ GeV and
tan$\beta \sim 2$) or the gluino mass and the squark mass are small
(around 100 GeV and 200 GeV respectively) and tan$\beta$ is also small.
While all other contributions suppressed in the large tan$\beta$ limit,
the only contribution surviving in this limit is the contribution
coming from exchanging neutrilino and down-type squarks and
the contribution can become important only in a very narrow region of
down-type squark mass in the low mass spectrum case\cite{ck}. In the 
following we limit ourself to discuss CP violation for $B^0$ and $\bar{B}^0$
decays far away from these regions, i.e., in the regions with large tan$\beta$
and relatively heavier down-type squark mass. Therefore, to a good 
approximation we can take the mixing to be that in SM, i.e.,
Eq.(\ref{mix}). Thus we can use Eqs. (\ref{CPT1}), (\ref{CPT2}), (\ref{CP1}) and (\ref{CP2})
in the numerical analysis. The explicit expressions of the Wilson coefficients $C_{10}, C_{Q_1}
, C_{Q_2}$ in a CP softly broken 2HDM and MSSM can be found in Refs.\cite{hz,hy}.

For a CP softly broken 2HDM\cite{hz}, the CP violation is depicted
by the phase of vacuum $\xi_H$ (i.e., $\xi$ in Ref. ~\cite{hz}). 
In Fig. 2 we give the plots of $A^2_{CP}$ for $B^0_d$ as functions of $\xi_H$
varying between $[0,2 \pi]$. Other parameters describing the model are 
chosen as $M_{H1}=120 ~GeV$, $M_{H2}=M_{H^\pm}=200 ~GeV$, tan$\beta=50$ for
which the experimental constraints of $K-\bar{K}$ and $B-\bar{B}$ mixing
, $\Gamma(b \to s \gamma)$, $\Gamma(b \to c \tau \bar{\nu}_\tau)$
and $R_b$ are well satisfied. The constraints
of electric dipole moments (EDMs) of the electron and neutron are also
satisfied except for $\xi_H=\frac{\pi}{4}$\cite{hz}. 
One may find that the CP asymmetry can be as large as $20 \%$ in vast
of the range of $\xi_H$ and can even reach $50 \%$.
For $B^0_s$, the dependence of the CP asymmetry on $\xi_H$ is similar to that for $B^0_d$
and the CP asymmetry can reach $3 \%$. 

For generic SUSY models, the constraints from the EDMs of the electron and
neutron on the CP violating phases have been analyzed
by many authors\cite{in,hl}. The scenario with large tan$\beta$,
which we are interested in here, have been discussed in our previous
papers\cite{hl}. The constraint of $\bsg$ has also been presented
there. In the case of low mass spectrum (the lighter
stop of order $200$ GeV and chargino masses less than $200$ GeV), $C_{Q1}$
and $C_{Q2}$ are constrained by the $\bsg$ decay,
because $C_{Q_i}$ and $C_7$ (the branching ratio of $b\rightarrow s \gamma$ is determined
by $|C_7|^2$ ) both receive most important SUSY contributions from
exchanging top squark. An interesting case happens when the SM contribution 
to $C_7$ is completely canceled by the real part of SUSY contributions and 
a considerable imaginary part is left\cite{hl} ( so that the
constraint on $C_7$ is satisfied ) if tan$\beta$ is large enough (say, $\geq 30$). 
$C_{Q1}$ and $C_{Q2}$, in this case, exhibit phases about $\pm \pi/4$ and consequently 
the absolute value of CP asymmetries for $B^0_d$ can be significantly larger than $30 \%$. 
CP asymmetries for $B^0_s$ can also be $\pm 3 \%$
in this case. As pointed out in Ref.\cite{hl}, the above areas of parameter space are allowed
by the EDM constraints due to the cancellation among the various contributions to EDMs.
 For the case of high mass spectrum where the $\bsg$
constraint can be safely satisfied and the CP violating phases of
trilinear term, $A_t$, and $\mu$ can survive in almost all of their
parameter space after satisfying the constraints of electron and neutron EDMs, 
the magnitudes of $C_{Q1}$ and
$C_{Q2}$ are also suppressed by the mass spectrum and CP asymmetries
can exhibit the correlation depicted in Fig. 1. But for this scenario
the branching ratio of the decay would not be enhanced large enough,
so it is less interesting.
In the supergravity(SUGRA) model there is another feature which
would have important phenomenological implications, i.e., because
electroweak (EW) symmetry is broken spontaneously by the radiative breaking mechanism 
the masses of the two heavier neutral Higgs bosons are of the same order.
Hence in general there is a large cancellation happened in the numerator
of Eq. (\ref{rate}) in SUGRA models. The consequence of it is that for
$B^0_d$, $A^1_{CP}$ is greatly suppressed(see Eq. (\ref{CP1})) even in the case
of low mass spectrum and the two CP asymmetries for $B^0_s$ are both small
($< 10^{-2}$). 

With the branching ratios $Br(B^0_q \to l^+_L l^-_L)$ and
$Br(B^0_q \to l^+_R l^-_R)$ given respectively by
\beq
  C_{B^0_q}\times \bigg[ (1-4{\hat m_l^2})|C_{Q_1}|^2
   + |C_{Q_2}+2 {\hat m_l} C_9|^2 -2 \sqrt{1-4 {\hat m_l^2}}
   [C_{Q1}^* \times(C_{Q2}2 +{\hat m_l} C_9)+ h.c.]
\bigg] \nnb \\
\eeq
and
\beq
 C_{B^0_q}\times \bigg[ (1-4{\hat m_l^2}) |C_{Q_1}|^2
   + |C_{Q_2}+2 {\hat m_l} C_9|^2  +2 \sqrt{1-4 {\hat m_l^2}}
  [C_{Q1}^* \times (C_{Q2}+2 {\hat m_l} C_9)+ h.c.]
\bigg] \nnb
\eeq
where
\beq
C_{B^0_q}=\frac{G_F^2 \alpha_{EM}^2}{64 \pi^3} m_{B^0_q}^3
\tau_{B^0_q} f^2_{B^0_q} |\lambda_t|^2 \sqrt{1 - 4 {\hat m_l^2}}, \nnb
\eeq
we calculate the events  $N^i_q$ (i=1,2) needed for observing $A_{CP}^i$
at 1$\sigma$ in the areas of parameter space in which $A_{CP}^i$ and the branching ratios
both have large values and all experimental constraints are satisfied. For l=$\mu$, they are order 
of $10^8$ and $10^9$ for $B_d^0$ and $B_s^0$ respectively
in 2HDM with CP violating phase and tan$\beta$=50 or in SUSY with tan$\beta$=30 as well as 
sparticle masses in a reasonable range. Therefore,
 $10^{10}$ ($10^{11}$) $B_d$ ($B_s$)
per year, which is in the designed range in the future B factors with $10^8$-
$10^{12}$ B hadrons per year~\cite{bs}, is needed in order to observe the CP asymmetry in $B\rightarrow
\mu^+\mu^-$ with good accuracy. For l=$\tau$, the events $N^i_q$ are order of $10^6$ and $10^7$ for $B_d^0$ and 
$B_s^0$ respectively in 2HDM with CP violating phase and tan$\beta$=50 or in SUSY with tan$\beta$=30 
as well as  sparticle masses in a reasonable range. 
Assuming a total of $5\times 10^8 (10^9)$ $B_d\bar{B_d}$ ($B_s\bar{B_s}$) decays, one can expect to observe
$\sim 100 $ identified $B_q\rightarrow \tau^+\tau^-$ events, permitting a test of the 
predicted CP asymmetry with good accuracy. 

As discussed above, we need to seperate the final state $l^+_L l^-_L$ from $l^+_R l^-_R$  in order to measure
CP asymmetry. For l=$\tau$, the polarization analysis is straightforward. However, detecting tau's is difficult experimentally.
For l=$\mu$, in principle one can separate the final state $\mu^+_L \mu^-_L$ from $\mu^+_R \mu^-_R$ by measuring
the energy spectra of the electron from muon decay\cite{pes}. A $\mu_L$ will decay to an
energetic $e_L$, which must go forward to carry the muon spin, and a less energetic pair
of neutrino and antineutrino because the electron is always left-handed\footnote{In the present 
case it is quite a good approximation to ignore the mass of electron.} and the energy-momentum 
and angular momentum are conserved. Due to the same reason, for  $\mu_R$, the relative energies of
electron and a pair of neutrino and antineutrino are roughly reversed. Therefore, the energy spectra of the 
electron from muon decay  is a powerful $\mu$ spin analyzer. However, in practice muons never decay in a 4$\pi$
detector because the lifetime of a muon is long ( c$\tau$=659 m). A possible way to measure a polarized muon decay is to build 
special detectors which can make muons lose its energy but keep polarization so that the polarized muon decays can be measured. 
 
In summary, we have analyzed the CP violation  in decays $B^0_q\rightarrow
l^+l^-$ (q=d,s). While there is no direct CP violation, there might be
mixed CP violation in the process 
\begin{equation}
B^0\rightarrow\bar{B}^0\rightarrow f~~~~~~~vs.~~~~~~~~\bar{B}^0
\rightarrow B^0\rightarrow \bar{f}. \nnb\\
\end{equation}
It is shown that in the approximation of $|\frac{q}{p}|$=1 the CP violation vanishes in SM. 
If including the correction of order of $10^{-2}$ to $|\frac{q}{p}|$=1, CP violation is smaller than
or equal to O($10^{-3}$) which is unobservably small. 
In a 2HDM with CP violating phases and MSSM the CP asymmetries depend on the 
parameters of models and can be as large as $40\%$ for $B^0_d$ and $3\%$ for $B^0_s$. 
Therefore, an observation of CP asymmetry in the decays 
$B^0_q \to l^+l^- (q=d,s)$ would unambiguously signal the existence of new physics.

We are grateful to Abdus Salam ICTP where part of the work was done for the support 
of ICTP Associateship Programme. We thank Qi-Shu Yan and Shou-Hua Zhu for discussions.  
C.-S. Huang would like to thank CSSM at Adelaide University where the letter
was revised for warm hospitality.
W. Liao would like to thank P. Ball for helpful communication. This research was supported
in part by the National Nature Science Foundation of China.

Figure Captions\\

Fig. 1 The correlation between CP asymmetries for $B^0_d$ and $B^0_s$.

Fig. 2 $A^2_{CP}$ for $B^0_d$ versus the CP violating phase $\xi_H$ in 
2HDM.

\begin{figure}
\begin{minipage}[t]{6.1 in}
\vskip+5cm     \epsfig{file=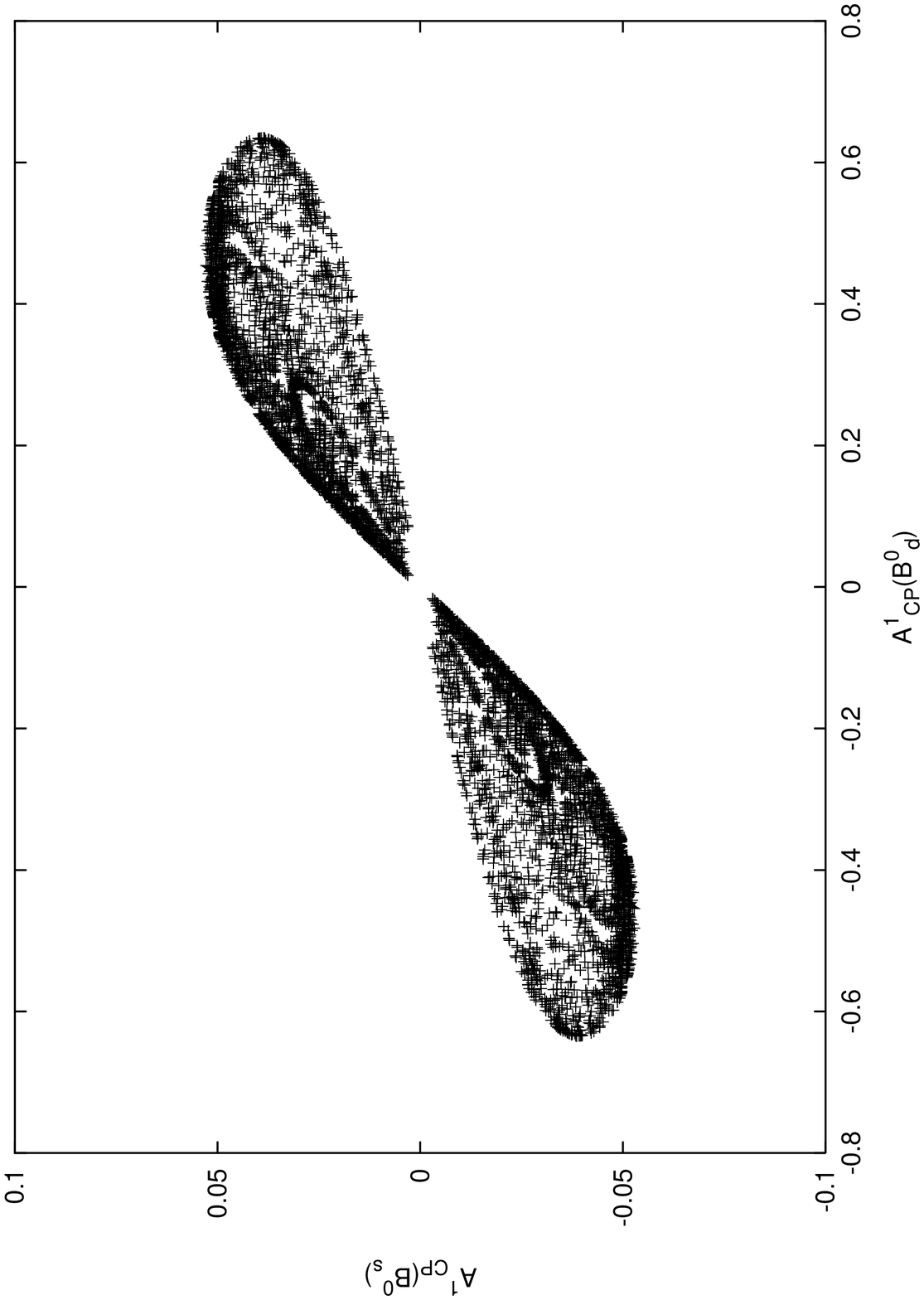, width=8.1in}
\vskip+1cm \hspace{8cm}  \caption{}
\end{minipage}
\end{figure}

\begin{figure}
\begin{minipage}[t]{5.1 in}
\vskip+5cm     \epsfig{file=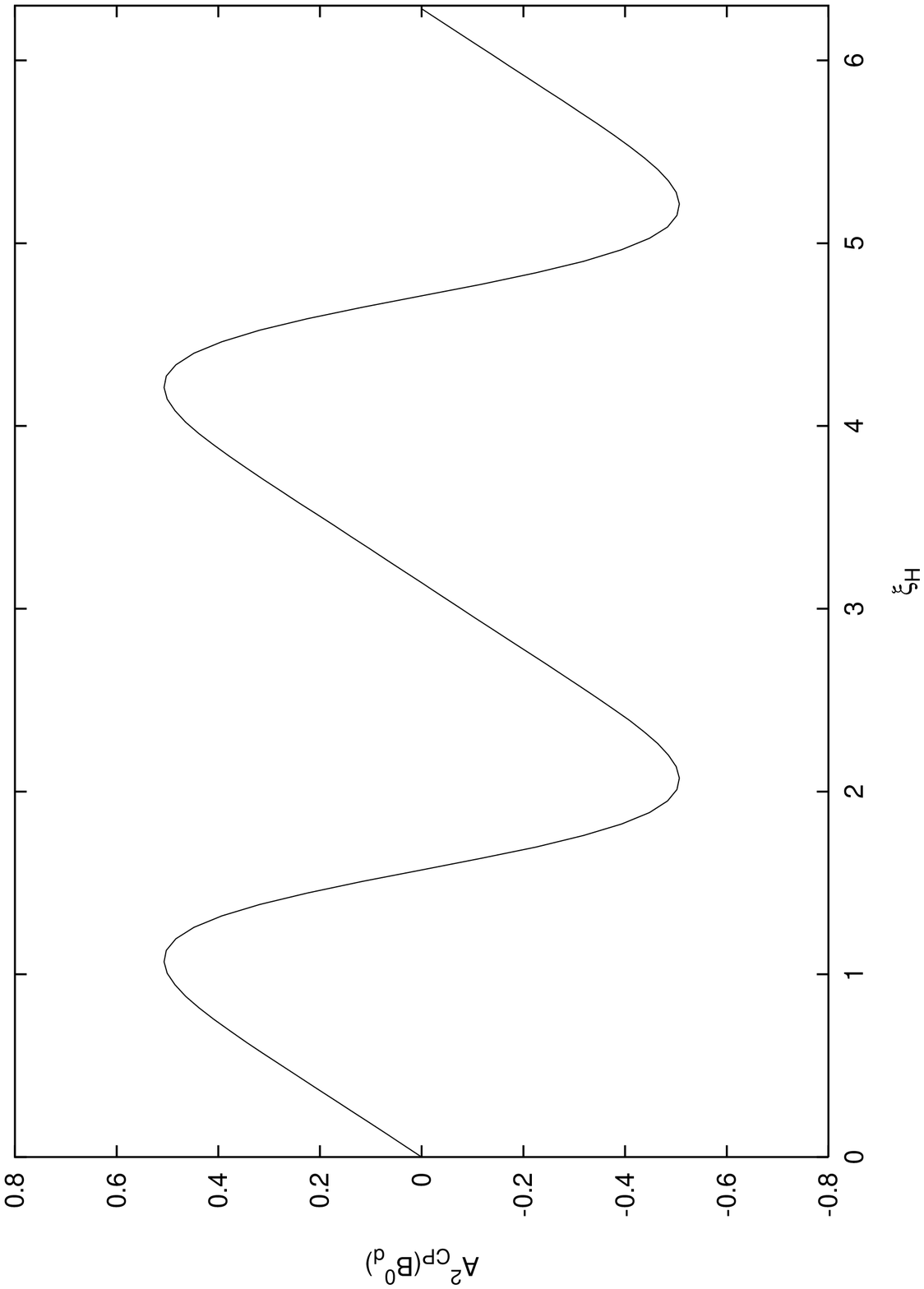, width=8.1in}
\vskip+1cm \hspace{8cm}  \caption{}
\end{minipage}
\end{figure}


\begin{thebibliography}{99}
\bibitem{hnv}
 X.G. He, T.D. Nguyen and R.R. Volkas, Phys. Rev. {\bf D38} (1988) 814;
W. Skiba and J. Kalinowski, Nucl. Phys. {\bf B404} (1993) 3;
D. Atwood, I. Reina and A. Soni, Phys. Rev. D{\bf 55}(1997) 3156;
H.E. Logan and U. Nierste, hep-ph/0004139; 
K.S. Babu and C. Kolda, Phys. Rev. Lett. {\bf 84}(2000) 228. 
\bibitem{hy}Y.-B. Dai, C.-S. Huang and H.-W. Huang, Phys. Lett. {\bf B390} (1997) 257;
C-S Huang and Q-S Yan, Phys. Lett. {\bf B442} (1998) 209;
C-S Huang, W. Liao and Q-S Yan, Phys. Rev.{\bf D59} (1999) 011701.
\bibitem{hlyz} C-S Huang, LIAO Wei, Q-S Yan and S-H Zhu, hep-ph/0006250;  C.-S. Huang, hep-ph/0009149.
\bibitem{cs}P.H. Chankowski, L. Slawianowska, hep-ph/0008046.
\bibitem{data} PDG, Review of Particle Data 2000.
\bibitem{Osaka}
Plenary talks presented by 
D. Hitlin (BaBar Collaboration) and H. Aihara (Belle Collaboration)
at ICHEP2000, Osaka, Japan, 31 July 2000 (to appear in the 
Proceedings).
\bibitem{cdf}Based the results given in the following references: OPAL Collaboration, K. Ackerstaff
et al., Eur. Phys. Jour. {\bf C5} (1998) 379; CDF Collaboration, T. Affolder et al., Phys. Rev.
{\bf D61} (2000) 072005.
\bibitem{bf}N. Ellis (LHC), at IV International 
Conference on Hyperons, Charm and Beauty Hadrons,
Valencia, Spain, 29 June 2000 (to appear in the proceedings). 
\bibitem{ks} F. Kr$\ddot{\textrm u}$ger and L.M. Sehgal, 
Phys.Lett. B380 (1996) 199.
\bibitem{hl} C-S Huang and LIAO Wei, Phys.Rev. D61 (2000) 116002;
Phys.Rev. D62 (2000) 016008.
\bibitem{gron} M. Gronau, Phys. Rev. Lett. {\bf 63}(1989) 1451.
\bibitem{wise}{B. Grinstein, M.J. Savage and M.B. Wise, Nucl. Phys. {\bf B319}
(1989)271.}
\bibitem{buras}G. Buchalla, A.J. Buras and M.E. Lauthenbache, Rev. Mod. Phys. {\
bf 68} (1996)1125.
\bibitem{75}A. Pais and S. Treiman, Phys. Rev. {\bf D12} (1975) 2744.
\bibitem{mas}S. Bertolini, F. borzumati, A. Masieso and G. Ridolfi, Nucl. Phys. 
{\bf B353} (1991) 591.
\bibitem{nir} Y. Nir, hep-ph/9911321.
\bibitem{bur}A.J. Buras, Probing the Standard Model of Particle Interactions,
Edited by F. David and R. Gupta, Elsevier Science(hep-ph/9806471).
\bibitem{ros}I. Dunietz and J.L. Rosner, Phys. Rev. {\bf D34} (1986) 1404.
\bibitem{ck} G. Couture, H. K$\ddot{\textrm o}$nig, Z. Phys. C{\bf 69}(1996)
 499; C{\bf 72}(1996) 327.
\bibitem{hz} C.-S. Huang, Z.-H. Zhu, Phys.Rev. D61 (2000) 015011; Erratum-ibid. D61 (2000) 119903.
\bibitem{in} T.Ibrahim and P.Nath, Phys. Lett. {\bf B418} 98 (1998), Phys.Rev. D
{\bf 57}, 478(1998), (E) ibid, {\bf D58}, 019901(1998), Phys. Rev. {\bf D58},111301(1998),
 hep-ph/9910553; M.Brhlik,G.J.Good,and G.L.Kane, Phys. Rev. {\bf D59},11504(1999);
D. Chang, W.-Y. Keung, A. Pilaftsis, Phys. Rev. Lett. {\bf 82} (1999) 900.
\bibitem{pes}A similar analysis was presented for seperating $t_L$ from $t_R$. See, C.R. Schmidt and M.E. Peskin, Phys. Rev. Lett. {\bf 69} (1992) 410.
\bibitem{bs}Belle Progress Report, Belle Collaboration, KEK- PROGRESS-REPORT-97-1 (1997);
Status of the BaBar Detector, BaBar Collaboration, SLAC-PUB-7951, presented at 29th International
Conference on High Energy Physics, Vancouver, Canada, 1998.
\end{thebibliography}
\end{document}